\documentclass[conference,10pt]{IEEEtran}
\ifCLASSINFOpdf
\else
\fi

\usepackage{stfloats}
\usepackage{makeidx}  

\usepackage{graphicx}
\usepackage{verbatim}
\usepackage{amsmath}
\usepackage{algorithm}
\usepackage{algpseudocode}
\usepackage{algorithmicx}
\usepackage{tabularx}
\usepackage{multirow}
\usepackage{hhline}
\usepackage{color}

\usepackage[nolist]{acronym}
\begin{acronym}
	\acro{AAA}{Authentication, Authorization and Accounting}
	\acro{AP}{Access Point}
	\acro{API}{Application Programming Interface}
	\acro{BF}{Bloom Filter}
	\acro{CBF}{Counting Bloom Filter}
	\acro{BP}{Baseline Pseudonym}
	\acro{BS}{Base Station} 
    \acro{BSM}{Basic Safety Message}
    \acro{BYOD}{Bring Your Own Device}
	\acro{C2C-CC}{Car2Car Communication Consortium}
	\acro{CA}{Certification Authority}
	\acrodefplural{CA}{Certification Authorities}
	\acro{CN}{Common Name}
	\acro{CAM}{Cooperative Awareness Message}
	\acro{CIA}{Confidentiality, Integrity and Availability}
	\acro{CRL}{Certificate Revocation List}
	\acro{CDN}{Content Delivery Network}
	\acro{CSR}{Certificate Signing Requests}
	\acro{DAA}{Direct Anonymous Attestation}
	\acro{DDoS}{Distributed Denial of Service}
	\acro{DDH}{Decisional Diffie-Helman}
	\acro{DENM}{Decentralized Environmental Notification Message}
	\acro{DHT}{Distributed Hash Table}
	\acro{DoS}{Denial of Service}
	\acro{DPA}{Data Protection Agency}
	\acro{ECU}{Electronic Control Unit}
	\acro{ETSI}{European Telecommunications Standards Institute}
	\acro{ECDSA}{Elliptic Curve Digital Signature Algorithm}
	\acro{ECC}{Elliptic Curve Cryptography}
	\acro{EVITA}{E-safety Vehicle Intrusion protected Applications}
	\acro{FCFS}{First-Come First-Served}
	\acro{FOT}{Field Operational Test}
	\acro{FPGA}{Field-Programmable Gate Array}
	\acro{FPL}{Fake Pseudonym List}
	\acro{GN}{GeoNetworking}
	\acro{GS}{Group Signatures}
	\acro{GM}{Group Manager}
	\acro{GBA}{Generic Bootstrapping Architecture}
	\acro{GPA}{Global Passive Adversary}
	\acro{GUI}{Graphic User Interface}
	\acro{HP}{Hybrid Pseudonym}
	\acro{HSM}{Hardware Security Module}
	\acro{HTTP}{Hypertext Transfer Protocol}
	\acro{IEEE}{Institute of Electrical and Electronics Engineers}
	\acro{IoT}{Internet of Things}
	\acro{ITS}{Intelligent Transport Systems}
	\acro{IT}{Information Technologies}
	\acro{IMSI}{International Mobile Subscriber Identity}
	\acro{IMEI}{International Mobile Station Equipment Identity}
	\acro{IdP}{Identity Provider}
	\acro{ISP}{Internet Service Provider}
	\acro{LEA}{Law Enforcement Agency}
	\acro{LTC}{Long-Term Certificate}
	\acro{LTCA}{Long-Term \acl{CA}}
	\acrodefplural{LTCA}{Long-Term \aclp{CA}}
	\acro{LDAP}{Lightweight Directory Access Protocol}
	\acro{LTE}{Long Term Evolution}
	\acro{LBS}{Location-based Service}
	\acro{MAC}{Message Authentication Code}
	\acro{MCA}{Message \ac{CA}}
	\acro{MEA}{Misbehavior Evaluation Authority}
	\acro{OBU}{On-Board Unit}
	\acro{OCSP}{Online Certificate Status Protocol}
	\acro{OSN}{Online Social Network}
	\acro{PC}{Pseudonymous Certificate}
	\acro{PCA}{Pseudonymous \acl{CA}}
	\acrodefplural{PCA}{Pseudonymous \aclp{CA}}
	\acro{PDP}{Policy Decision Point}
	\acro{PEP}{Policy Enforcement Point}
	\acro{PIR}{Private Information Retrieval}
	
	\acro{PKI}{Public-Key Infrastructure}
	\acro{POI}{Point of Interest}
	\acro{PRECIOSA}{Privacy Enabled Capability in Co-operative Systems and Safety Applications}
	\acro{PRESERVE}{Preparing Secure Vehicle-to-X Communication Systems}
	\acro{P2P}{peer-to-peer}
	\acro{RA}{Resolution Authority}
	\acro{REST}{Representational State Transfer}
	\acro{RBAC}{Role Based Access Control}
	\acro{RCA}{Root \acl{CA}}
	\acro{RSU}{Roadside Unit}
	\acro{SAML}{Security Assertion Markup Language}
	\acro{SCORE@F}{Système COopératif Routier Expérimental Français}
	\acro{SeVeCom}{Secure Vehicle Communication}
	\acro{SIS}{Security Infrastructure Server}
	\acro{SP}{Service Provider}
	\acro{SSO}{Single-Sign-On}
	\acro{SoA}{Service-oriented-Approach}
	\acro{SOAP}{Simple Object Access Protocol}
	\acro{SAS}{Sample Aggregation Service}
	\acro{TS}{Task Service}
	\acro{TLS}{Transport Layer Security}
	\acro{TPM}{Trusted Platform Module}
	\acro{TTP}{Trusted Third Party}
	\acro{TVR}{Ticket Validation Repository}
	\acro{URI}{Uniform Resource Identifier}
	\acro{VANET}{Vehicular Ad-hoc Network}
	\acro{V2I}{Vehicle-to-Infrastructure}
	\acro{V2V}{Vehicle-to-Vehicle}
	\acro{V2X}{\ac{V2V} and/or \ac{V2I}}
	\acro{VC}{Vehicular Communication}
	\acro{VM}{Virtual Machine}
	\acro{VSN}{Vehicular Social Network}
	\acro{VSS}{\ac{VC} Security Subsystem}
	\acro{WAVE}{Wireless Access in Vehicular Environments}
	\acro{WSDL}{Web Services Discovery Language}
	\acro{W3C}{World Wide Web Consortium}
	\acro{VANET}{Vehicular Ad-hoc Network}
	\acro{VPKI}{Vehicular Public-Key Infrastructure}
	\acro{WS}{Web Service}
	\acro{WoT}{Web of Trust}
	\acro{WSACA}{\ac{WAVE} Service Advertisement \ac{CA}}
	\acro{XML}{Extensible Markup Language}
	\acro{XACML}{eXtensible Access Control Markup Language}
	\acro{3G}{3rd Generation}
\end{acronym}

\usepackage{algpseudocode}
\usepackage{varwidth}
\usepackage{subcaption}

\usepackage{algorithm}
\PassOptionsToPackage{noend}{algpseudocode}
\usepackage{algpseudocode}
\usepackage{tikz}
\usetikzlibrary{shapes,arrows}

\errorcontextlines\maxdimen

\makeatletter
\newcommand*{\algrule}[1][\algorithmicindent]{\makebox[#1][l]{\hspace*{.5em}\vrule height .75\baselineskip depth .25\baselineskip}}%

\newcount\ALG@printindent@tempcnta
\def\ALG@printindent{%
	\ifnum \theALG@nested>0
	\ifx\ALG@text\ALG@x@notext
	\addvspace{-3pt}
	\else
	\unskip
	\ALG@printindent@tempcnta=1
	\loop
	\algrule[\csname ALG@ind@\the\ALG@printindent@tempcnta\endcsname]%
	\advance \ALG@printindent@tempcnta 1
	\ifnum \ALG@printindent@tempcnta<\numexpr\theALG@nested+1\relax
	\repeat
	\fi
	\fi
}%
\usepackage{etoolbox}
\patchcmd{\ALG@doentity}{\noindent\hskip\ALG@tlm}{\ALG@printindent}{}{\errmessage{failed to patch}}
\makeatother

\begin{document}
	%
	\title{Proactive Certificate Validation for VANETs}

	\author{\IEEEauthorblockN{Hongyu Jin and Panos Papadimitratos}
	\IEEEauthorblockA{Networked Systems Security Group, KTH Royal Institute of Technology, Sweden \\
		\emph{\{hongyuj, papadim\}}@kth.se
		}}


	\maketitle
	
	\begin{abstract}
		Security and privacy in \acp{VANET} mandates use of short-lived credentials (pseudonyms) and cryptographic key pairs. This implies significant computational overhead for vehicles, needing to validate often numerous such pseudonyms within a short period. To alleviate such a bottleneck that could even place vehicle safety at risk, we propose a proactive pseudonym validation approach based on \acp{BF}. We show that our scheme could liberate computational resources for other (safety- and time-critical) operations with reasonable communication overhead without compromising security and privacy.
	\end{abstract}
	

	%
	\IEEEpeerreviewmaketitle

\section{Introduction}

\ac{V2V} communication improves road safety and traffic efficiency with safety beacons, broadcasted at a high rate to provide cooperative awareness. Short-term credentials, i.e., pseudonyms obtained from the \ac{VPKI} through protocols as in, e.g.,~\cite{c2c, whyte2013security,khodaei2014towards}, are used for message (beacon) authentication and integrity while protecting user privacy. Pseudonyms with overlapping or non-overlapping lifetimes can be preloaded, for a long period (e.g., 1 year) or be requested on-demand (e.g., on a daily basis). In a multi-domain \ac{VC} system, pseudonyms in a domain are generally issued by the \ac{PCA} dedicated to that domain, and a vehicle that wishes to enter another (foreign) domain should request pseudonyms from the corresponding \ac{PCA}. For ease of explanation, we assume the domains are separated geographically in the rest of the paper.

Pseudonyms are changed over time for message unlinkablity. Due to mobility of vehicles, the neighborhood of a vehicle can be volatile, thus, having new pseudonyms received practically continuously. The challenge is that all such digitally signed new pseudonyms must be validated in order to verify messages. Certificate omission~\cite{calandriello2011performance}, and optimistic or probabilistic message validations~\cite{jin2015scaling,kargl2015wireless,lyu2016pba} have been proposed, but they do not reduce pseudonym validation overhead. In some situations, a vehicle could receive a very large number of new pseudonyms within a short period (e.g., around a mix-zone~\cite{freudiger2007mix}, where all vehicles would change their pseudonyms).



We propose a \acf{BF} based pseudonym validation scheme. Instead of verifying the \ac{PCA} signature for each and every pseudonym, the pseudonyms are validated through a \ac{BF} published by the \ac{PCA}, which includes all pseudonyms valid within a protocol selectable period. Once the \ac{BF} is verified and stored, a vehicle can efficiently validate the pseudonyms based on cheap hash computations with reasonably low false positive rate.

We require that all the pseudonyms are still signed by the \ac{PCA} and the messages be signed under the pseudonyms. This ensures that a fallback approach (i.e., \ac{PCA} signature verification on each and every pseudonym) can be invoked when suspicious behavior is detected. We show that our scheme could reduce computational overhead. Although an attacker could launch a brute false attack targeting the false positive rate of the \ac{BF} (attempting to inject messages signed under fictitious pseudonyms), we show that such an attack is expensive and could cause minimal harm to the system.

In the rest of the paper, we describe the adversary model (Sec.~\ref{sec:model}), present our pseudonym validation scheme inspired by~\cite{ren2009multi} (Sec.~\ref{sec:scheme}), provide a security and privacy analysis (Sec.~\ref{sec:analysis}), and a preliminary evaluation of our scheme (Sec.~\ref{sec:evaluation}) before some concluding remarks (Sec.~\ref{sec:conclusion}).



\section{Adversary Model}
\label{sec:model}


We consider (external or internal) adversaries that attempt to insert false messages, without using a legitimate private/public key pair and the corresponding pseudonym in order to affect other vehicles. Such an attack could inject, for example, a false event (e.g., a non-existing accident). In addition, we consider adversaries interested in launching clogging \ac{DoS} attacks, i.e., sending out messages with fake signatures at a high rate, in order to consume resources of benign vehicles (and leave them with scarcer resources for processing legitimate, and potentially critical, messages).

Internal adversaries could threaten the network by sending out false information with valid signatures under valid pseudonyms. Misbehavior detection would then lead to their identification and eviction; this is orthogonal to this paper.
\section{Our Scheme}
\label{sec:scheme}

\subsection{Preliminaries}

\textbf{Counting Bloom Filter:} \acp{BF}~\cite{mitzenmacher2002compressed} are used for efficient membership checking in Internet applications. A \ac{BF} is built based on elements of a dataset, and the published \ac{BF} can be used for membership checking for a given element. Each element in the set is hashed with $k$ hash functions, while the output of each hash function is a position in an $m$-bit vector and these $k$ positions are set to 1. However, if any of them is already set to 1 upon a previous insertion, these bits are simply kept as 1 and ignored. For a membership checking, the element is hashed with $k$ hash functions, and the derived $k$ positions are compared with the \ac{BF}. If all $k$ positions are 1, then the element has passed the membership test. A \ac{BF} reduces spatial overhead at the expense of a false positive rate. An element not included in the original dataset could pass the \ac{BF} test if all $k$ positions for this element were set to 1 by other elements. For a standard \ac{BF} (the type we consider in our paper), $m$ and $k$ are chosen based on the number of the dataset elements, $n$, and the false positive rate to minimize spatial overhead, $m$~\cite{mitzenmacher2002compressed}.

A standard \ac{BF} supports insertions of new elements but no deletions: a bit in the \ac{BF} might be needed by multiple elements. A new \ac{BF} has to be built from scratch if elements are deleted. Counting \acp{BF}~\cite{mitzenmacher2002compressed} maintain a counter for each bit, indicating the times it was set to 1. Therefore, when an element is deleted, for each of its $k$ bits, the counter is decreased by 1. If a counter is decreased to 0, then the corresponding bit in the \ac{BF} is also set to 0. The size of a counter should be chosen properly~\cite{mitzenmacher2002compressed}.

\textbf{Compressed \ac{BF}-Delta:} Compressed \ac{BF}-deltas~\cite{mitzenmacher2002compressed} can be used to publish updates when a few of the \ac{BF} elements are changed (e.g., inserted or deleted). This provides an efficient way to publish differences (in terms of each bit value) between old and new \acp{BF} with minimum overhead.

\textbf{Note:} An \emph{alternative} to \acp{BF} could be a concatenation of hash values for elements in the dataset, published as a hash list. For membership checking, the hash value of the element is computed and searched in the hash list. However, for a large dataset, \acp{BF} are far superior in terms of spatial overhead~\cite{bfvshash}. Moreover, searching in a hash list requires $O(n)$ time complexity, while a \ac{BF}-based checks require $O(k)$ time complexity (with $k \ll n$, typically, for any sizeable dataset).

\subsection{Scheme Overview}

We propose an efficient pseudonym validation scheme based on \acp{BF}. The \ac{PCA} generates a counting \ac{BF} based on all the currently valid pseudonyms issued to vehicles. Compressed \ac{BF}-deltas are used to publish updates in case of insertions (e.g., new pseudonyms provided in response to recent requests) and deletions (e.g., revocation of pseudonyms). Vehicles can download the \ac{BF} (without the counters) from the \ac{PCA} once it is built, and download periodically newer versions (or deltas). Once the \ac{BF} is downloaded, vehicles could validate received pseudonyms with the \ac{BF}, at a processing cost that is a tiny fraction of that to validate a digital signature by the \ac{PCA}. If a pseudonym does not pass the \ac{BF} test, e.g., in the event recently issued pseudonyms are not yet included in the \ac{BF}, a receiving/validating vehicle can always choose the fallback approach (referred as the \emph{baseline} scheme in the rest of this paper): verify the \ac{PCA} signature on the pseudonym.

\subsection{Bloom Filter based Pseudonym Validation}

\begin{figure}[t]
	\centering
	\includegraphics[width=\columnwidth]{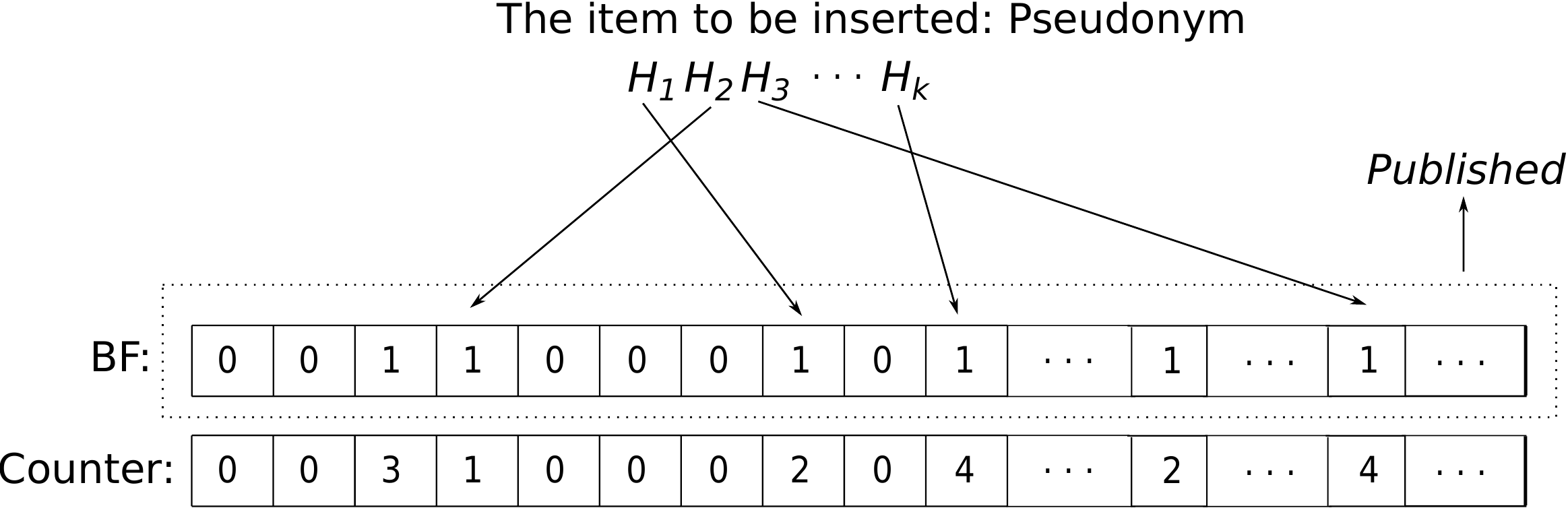}
	\caption{BF Construction with $k$ hash functions}
	\label{fig:bf_auth}
	\vspace{-1em}
\end{figure}

Without loss of generality, we assume the majority of vehicles (e.g., local vehicles) have been preloaded with pseudonyms for a period, $\Gamma$ (e.g., $24\,h$), thus covering $[t_{start}, t_{start} + \Gamma]$. We assume these pseudonyms are requested well in advance before $t_{start}$. The \ac{PCA} generates a \ac{BF} that includes the pseudonyms covering $[t_{start}, t_{start} + \Gamma]$. We do not dwell on the selection of $t_{start}$; e.g., a point during the night could be chosen, so that vehicles request pseudonyms and download the new \ac{BF} while parked.

Pseudonyms can either have overlapping (e.g., 100 pseudonyms for each vehicle, all valid for $24\,h$) or non-overlapping (e.g., 144 pseudonyms for each vehicle, each valid for $10\,min$) lifetimes. In the former case, an element in the \ac{BF} is the public key of a pseudonym; while for the latter case, an element is the combination of a public key and its corresponding lifetime. Fig.~\ref{fig:bf_auth} shows the construction of the \ac{BF} based on the pseudonyms. Although the \ac{PCA} maintains a counting \ac{BF}, only a standard \ac{BF} is published, because the counters are not necessary for pseudonym validation; counters are used to support insertions and deletions to the \ac{BF}. While a larger counter size results in higher storage overhead (for the \ac{PCA}), this does not affect the size of downloaded \ac{BF} (thus the communication overhead for the vehicles).

Vehicles that did not request pseudonyms from the \ac{PCA} before $t_{start}$ could request pseudonyms throughout the day. This can be, e.g., due to non-predictable trips or new vehicles joining from other domains. As these vehicles request pseudonyms from the \ac{PCA}, the \ac{BF} has to be updated to cover these new pseudonyms. A vehicle could update the \ac{BF} either proactively, when the vehicle is parked, or reactively, when it starts receiving a considerable amount (above a protocol-selectable threshold) of pseudonyms not included in the \ac{BF}. We use compressed \ac{BF}-deltas to minimize the communication overhead for updating the \ac{BF}.

As \ac{BF} exhibits a false positive rate, a fake pseudonym discovered by a brute force search (with very low probability if we choose $m$ appropriately) could be accepted even if it were not issued by the \ac{PCA} (see Sec.\ref{sec:analysis} for more details). This can be mitigated by applying a probabilistic verification even if a pseudonym passed the \ac{BF} test. If such a double-checked pseudonym is proven fake, it is reported to the \ac{VPKI} and published in a \ac{FPL}. An \ac{FPL} is a list of detected fake pseudonyms that could pass \ac{BF} tests.

\textbf{Validation process:} In order to validate a pseudonym, the receiver first tests the pseudonym against the currently available local version of the BF. If the \ac{BF} test is successful, the pseudonym is checked against the \ac{FPL}: the pseudonym is validated if it is not included in the \ac{FPL}. If the pseudonym did not pass the \ac{BF} test, the signature on the pseudonym has to be verified (i.e., the baseline scheme). To ensure resilience to clogging \ac{DoS}, the fraction of such baseline validation should be conservative and adaptive. In order to mitigate the effect of fake pseudonyms, for each pseudonym that passed \ac{BF} test and \ac{FPL} check, the receiver could verify probabilistically (with a low probability) the signature on the pseudonym. If this pseudonym cross-verification fails, the fake pseudonym is reported to the \ac{VPKI} and added to the \ac{FPL}.

\section{Security \& Privacy Analysis}
\label{sec:analysis}

\textbf{Non-repudiation, authentication and integrity:} We require that all valid pseudonyms and messages be signed by the \ac{PCA} and their senders respectively. Thus, our scheme does not affect the non-repudiation of the messages. The \ac{BF} and its deltas are authenticated and cannot be repudiated. If the received pseudonyms are not included in the local \ac{BF} or, in case, any suspicious actions are detected, vehicles can always validate pseudonyms using the baseline scheme.

\begin{figure}[t]
	\centering
	\includegraphics[width=0.95\columnwidth]{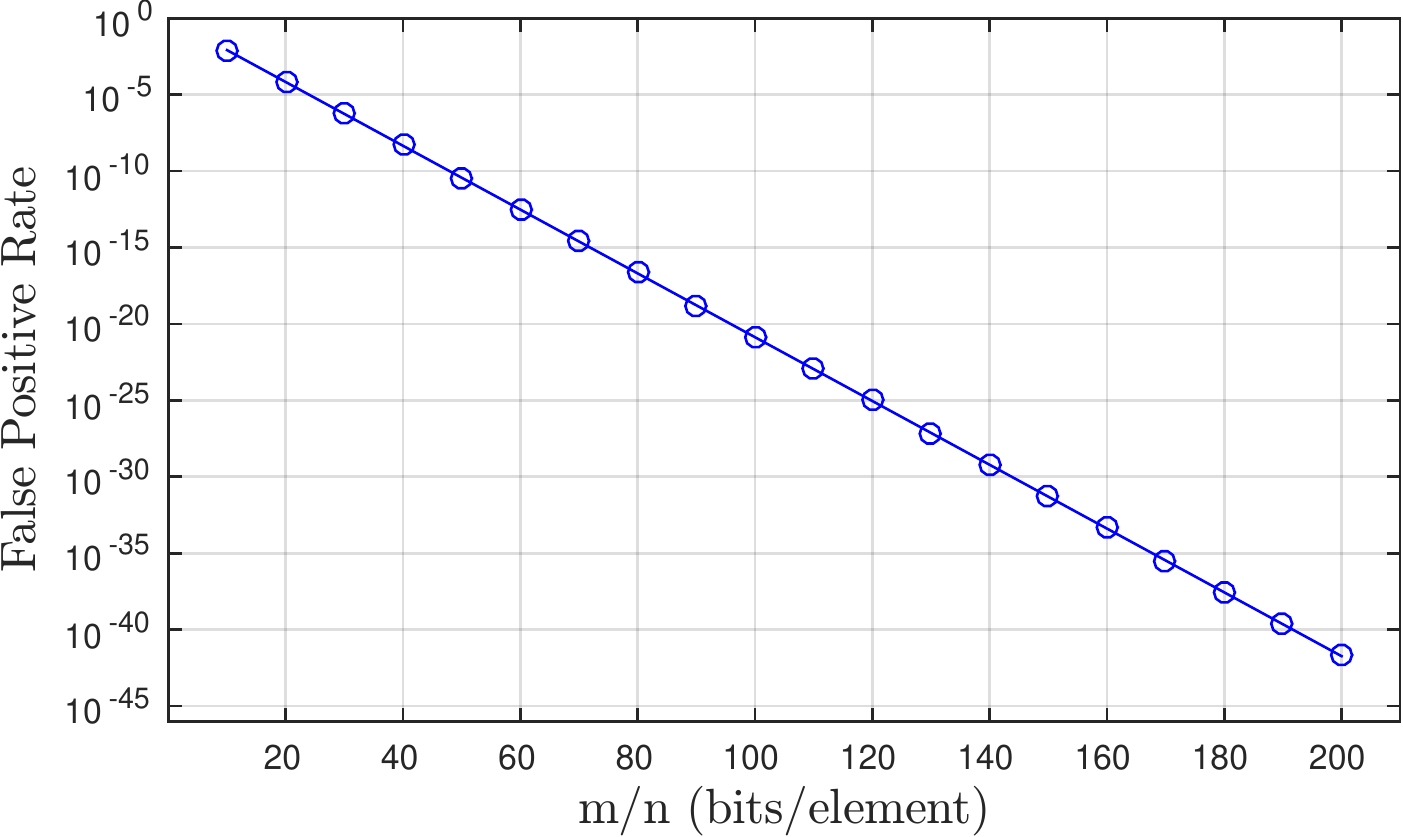}
	\caption{False positive rate as a function of $m/n$}
	\label{fig:false_positive}
\end{figure}

\textbf{Fake pseudonyms:} An adversary could target the false positive rate of a \ac{BF}, shown in Fig.~\ref{fig:false_positive} as a function of $m/n$ (bits per element)~\cite{mitzenmacher2002compressed}. For example, when $m/n = 80$ bits, the false positive rate is around $2 \times 10^{-17}$. Consider the case the pseudonyms have the same lifetime (e.g., valid for $24\,h$). On average, an attacker would have to generate around $10^{17}$ public/private key pairs to find a fake one passing the \ac{BF} test, each test needing $k$ hash computations. If the pseudonyms have non-overlapping lifetimes, a pair of public/private key could be tested with different lifetimes, thus less key pairs would be needed to find one passing the \ac{BF} test. However, such a fake pseudonym can only be used for a short period (i.e., within its lifetime). Moreover, probabilistic verification further shortens the period a fake pseudonym can be used before being detected. Last but not least, a limited number (depending on the processing power of the attacker) of fake pseudonyms would not significantly affect the system if vehicles with valid pseudonyms are the majority within the neighborhood. Overall, such a brute force attack is expensive, and it may have a negligible effect.


\textbf{Privacy of newly joining vehicles:} A \ac{PCA} would update its \ac{BF} as new vehicles join the domain (and request pseudonyms). However, this raises a privacy concern that pseudonyms of new vehicles could be easily linked. For example, if the \ac{BF} is updated for three new vehicles which appear in different parts of the domain; a global passive attacker could easily link the sets of pseudonyms that were not included in the old \ac{BF}. This can be mitigated by updating the \ac{BF} only when a considerable amount of new vehicles joins the domain, essentially, creating a larger anonymity set for those new comers.


\textbf{Thwarting clogging \ac{DoS}:} No invalid pseudonyms (or messages) would be accepted by the baseline scheme. However, this makes clogging likely: an attacker could generate arbitrary strings as public keys and attach arbitrary strings as signatures, and broadcast at a high rate. This kind of attack is cheap, while consuming resources of benign vehicles to verify the fake signatures. Optimistic message verifications have been widely studied while some of them rely on short-term linkability of the messages signed under the same pseudonym. For example, \cite{lyu2016pba} proposes to use TESLA for the verification of following messages after a signature verification on the first message. However, it would be pointless to thwart this kind of attack by sacrificing linkability among the pseudonyms. Our scheme can efficiently thwart such an attack, while an attacker needs significant effort to find false positive pseudonyms as we discussed earlier. If increasing amount of pseudonyms with fake signatures (that do not pass \ac{BF} tests) are received; the fraction of CPU time assigned for verifying signatures on pseudonyms (including both new legitimate pseudonyms and randomly generated pseudonyms) can be reduced and vehicles should update their \acp{BF} in order to properly validate legitimate pseudonyms.


\section{Performance Evaluation}
\label{sec:evaluation}

\subsection{Communication Overhead}


Consider a \ac{BF} with a false positive rate of $10^{-20}$. If $n=14\,400\,000$ ($100\,000$ vehicles equipped with $144$ pseudonyms valid for $24\,h$ each), $m \approx 164.5$ $Mbytes$ ($m/n \approx 96$ bits, see Fig.~\ref{fig:false_positive}). The original \ac{BF} can be downloaded before a trip starts, which takes, e.g., around $1\,min$ with a bandwidth of $20$ $Mbps$. The size of the \ac{BF} is acceptable considering the volumes that could be provided by off-the-shelf hard-drives nowadays.



The use of compressed \ac{BF}-delta~\cite{mitzenmacher2002compressed} decreases the communication overhead to update the \ac{BF}.  The compression rate can be computed, with $q = p(1-p^f)$:
\begin{align}
\text{\emph{Compression Rate }} = -q \log_{2}q - (1 - q) \log_{2}(1-q), \label{eq:compress_rate}
\end{align}
where $p$ is the probability that a bit in the original \ac{BF} is $1$, and $q$ is the probability that a bit in the \ac{BF}-delta is $1$ after a fraction, $f$, of pseudonyms are added to the original \ac{BF}. For a standard \ac{BF}, which we use in our scheme, $m/n$ and $k$ are chosen so that $p \approx 1/2$~\cite{mitzenmacher2002compressed}.

\begin{figure}[t]
	\centering
	\includegraphics[width=\columnwidth]{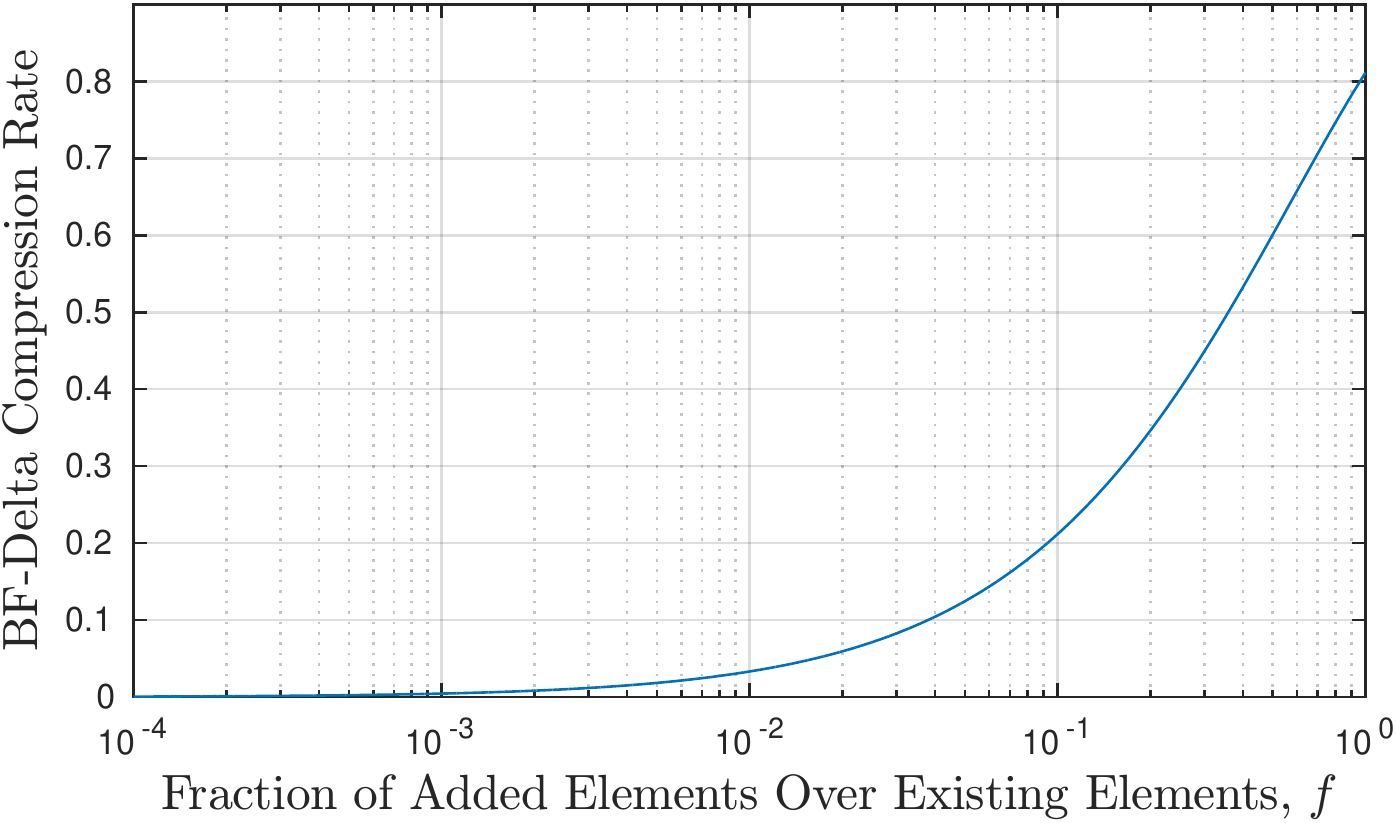}
	\caption{Compression rate of \ac{BF}-delta as a function of the fraction of added elements.}
	\label{fig:bf_delta}
\end{figure}

Fig.~\ref{fig:bf_delta} shows the compression rate of a \ac{BF}-delta as a function of $f$. If $144\,000$ new pseudonyms (for $1\,000$ new vehicles) were added, then the size of the compressed \ac{BF}-delta is around $6.2$ Mbytes (with a compression rate of 0.0045). For $10\,000$ new vehicles, it is $34.8$ Mbytes. We should note that the false positive rate of the \ac{BF} increases as pseudonyms are added, because more bits are set to $1$. The probability that a bit in the updated \ac{BF} is $1$ can be computed as: $p^\prime = p + (1-p)q$. For example, if $10\,000$ (i.e., $f = 0.1$) new vehicles would join everyday, $p^\prime = 0.6$ after adding the pseudonyms. Then, the false positive rate of the updated \ac{BF} is around $3.29 \times 10^{-20}$: a slight increase from $10^{-20}$. We refer the reader to~\cite{mitzenmacher2002compressed} for the false positive rate calculation.

\subsection{Computation Overhead}

In our scheme, validation of a pseudonym requires $k$ hash computations, which is much cheaper than a signature verification. Consider the following case: $N$ vehicles are within a vehicle's communication range and the neighborhood refresh/change (in terms of new pseudonyms) ratio is $c$ per second. Each vehicle broadcasts $\gamma$ beacons per second. We assume ECDSA-256 for both the pseudonyms and the \ac{PCA} certificate, and signature verification delay $\tau = 4$ $ms$ (a typical value from the literature~\cite{calandriello2011performance}). For simplicity, we assume the delay of a \ac{BF} test is $0$ $ms$ (in reality, it introduces a tiny delay for $k$ hash computations, which can be in the order of $\mu s$) and all the pseudonyms from $N$ neighbors are included in the \ac{BF}.

We consider a two-class $M/D/1$ queue for message verifications, as in~\cite{calandriello2011performance}. The first class includes messages signed under new pseudonyms and the second class includes messages signed under stored pseudonyms. The \emph{average system time}, $\bar{T}$ (total time in the queue until a message is verified), can be represented as:
\begin{align}
\bar{T} = \bar{S} + \frac{\lambda_1 S_1^2 + \lambda_2 S_2^2}{2(1-\rho)},
\end{align}
where $\bar{S}$ is the average service time, and $\lambda_i$ and $S_i$ are the arrival rate and the service time of $i$th class. We can derive that $\lambda_1=cN$, $\lambda_2=\gamma N - \lambda_1$ and $S_2=\tau$, while $S_1$ for the baseline and the \ac{BF}-based schemes are $2\tau$ and $\tau$ respectively. From \emph{Little's law}, we know that $\rho = \rho_1 + \rho_2$, $\rho_i = \lambda_i S_i$ and $\bar{S} = \frac{\rho}{\lambda_1 + \lambda_2}$.

\begin{figure}[t]
	\centering
	\includegraphics[width=0.96\columnwidth]{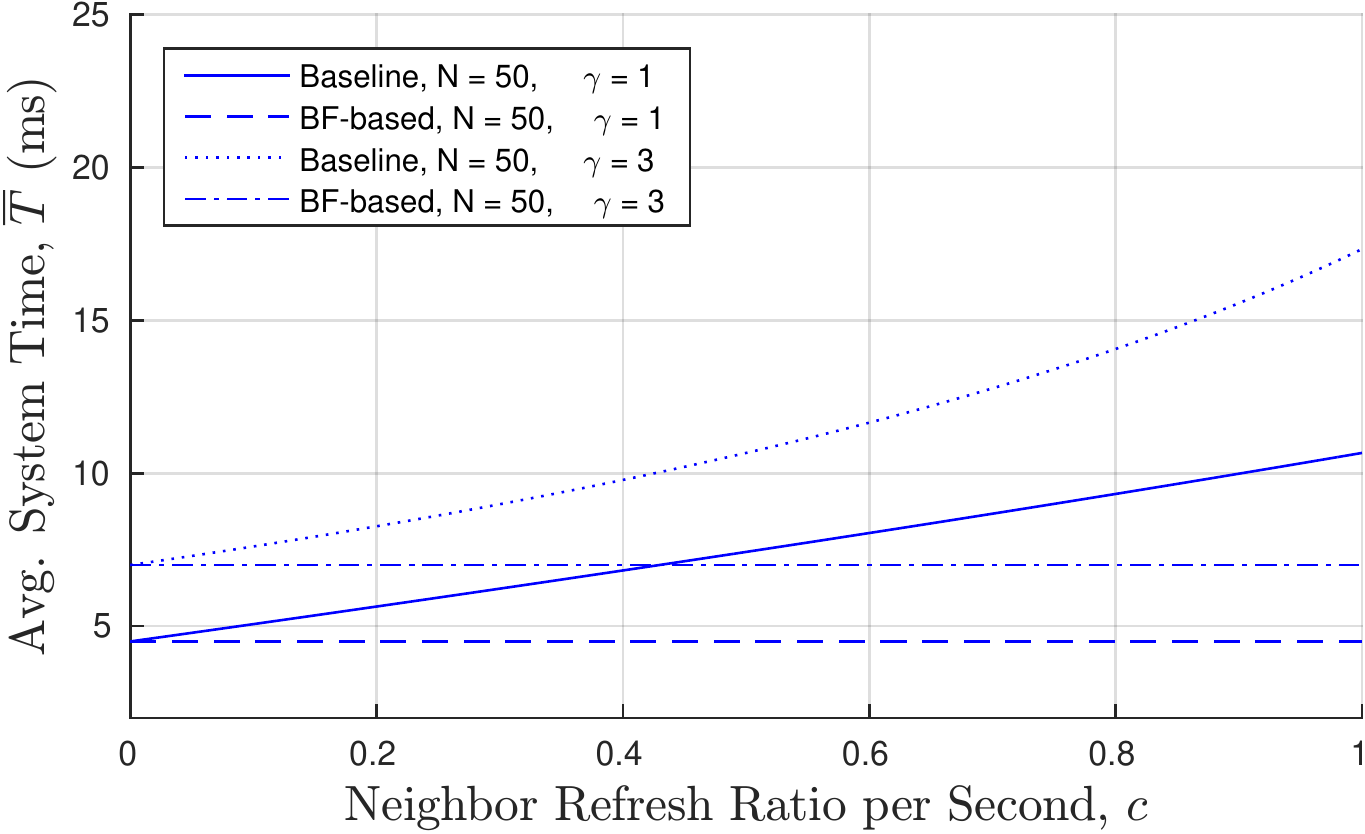}
	\caption{Average waiting time as a function of neighbor refresh ratio per second}
	\label{fig:system_time}
\end{figure}

Fig.~\ref{fig:system_time} shows $\bar{T}$ as a function of $c$ when $N=50$. As expected, $c$ does not affect $\bar{T}$ for the \ac{BF}-based scheme, because \ac{BF} tests introduce negligible delay. However, for the baseline scheme,  $\bar{T}$ increases as $c$ increases because more signature verifications are needed for the new pseudonyms. For example, when $\gamma = 3$ and $c = 0.6$, $\bar{T}$ with the baseline scheme is almost double of that with the \ac{BF}-based scheme (without proactive cross-verification).

\section{Conclusion and Future Work}
\label{sec:conclusion}

We presented a \ac{BF}-based pseudonym validation scheme. We showed that the \ac{BF} can be downloaded with acceptable overhead and it can be used to validate pseudonyms efficiently with a reasonably low false positive rate. Even though an attacker could launch a brute force attack on the \ac{BF}, this would be expensive and likely to cause minimal harm to the system.

In this paper, we consider and analyzed the scheme with one \ac{BF} per domain. The immediate extension is to generalize, e.g., with one \ac{BF} per \ac{PCA} (presuming multiple \acp{PCA} exist in a domain), and download the \acp{BF} from all the \acp{PCA} in the domain or even download \acp{BF} from \acp{PCA} in neighboring foreign domains (to facilitate ``roaming'').

\bibliographystyle{abbrv}
\bibliography{references.bib}
	
\end{document}